\pdfoutput=1

\documentclass[]{spie}  
\usepackage[]{graphicx}
\usepackage{siunitx}
\usepackage{graphics}

\title{The Cosmology Large Angular Scale Surveyor (CLASS): 38~GHz detector array of bolometric polarimeters} 

\author{John W. Appel\supit{a}, Aamir Ali\supit{a}, Mandana Amiri\supit{e}, Derek Araujo\supit{f}, 
Charles L. Bennett\supit{a}, Fletcher Boone\supit{a}, Manwei Chan\supit{a},
Hsiao-Mei Cho\supit{d}, David T. Chuss\supit{b}, Felipe Colazo\supit{b},
Erik Crowe\supit{b}, Kevin Denis\supit{b}, Rolando D\"unner\supit{c}, Joseph Eimer\supit{a},
Thomas Essinger-Hileman\supit{a}, Dominik Gothe\supit{a}, Mark Halpern\supit{e},
Kathleen Harrington\supit{a}, Gene Hilton\supit{d}, Gary F. Hinshaw\supit{e},
Caroline Huang\supit{a}, Kent Irwin\supit{g}, Glenn Jones\supit{f}, John Karakla\supit{a},
Alan J. Kogut\supit{b}, David Larson\supit{a}, Michele Limon\supit{f},
Lindsay Lowry\supit{a}, Tobias Marriage\supit{a}, Nicholas Mehrle\supit{a},
Amber D. Miller\supit{f}, Nathan Miller\supit{b}, Samuel H. Moseley\supit{b},
Giles Novak\supit{h}, Carl Reintsema\supit{d}, Karwan Rostem\supit{a}\supit{b},
Thomas Stevenson\supit{b}, Deborah Towner\supit{b}, Kongpop U-Yen\supit{b},
Emily Wagner\supit{a}, Duncan Watts\supit{a}, Edward Wollack\supit{b},
Zhilei Xu\supit{a}, and Lingzhen Zeng\supit{i}.
\skiplinehalf
\supit{a}Johns Hopkins University, Department of Physics and Astronomy, 3400 N. Charles Street, Baltimore, MD 21218, USA; \\
\supit{b}NASA Goddard Space Flight Center, 8800 Greenbelt Rd, Greenbelt, MD 20771, USA;\\
\supit{c}Pontificia Universidad Cat\'{o}lica de Chile, Instituto de Astrofisica, Av. Vicuna Mackena 4860, Santiago, Chile;\\
\supit{d}National Institute of Standards and Technology, 325 Broadway, Boulder, CO 80305, USA;\\
\supit{e}University of British Columbia, Department of Physics and Astronomy, 6224 Agricultural Road, Vancouver, BC V6T 1Z4, Canada;\\
\supit{f}Columbia University, Department of Physics,  704 Pupin Hall, MC 5255, 538, Manhattan, NY, USA;\\
\supit{g}Stanford University, Department of Physics, 382 Via Pueblo, Stanford, CA 94305, USA;\\
\supit{h}Northwestern University, Department of Physics and Astronomy, 2145 Sheridan Road, Evanston, IL 60208-3112, USA;\\
\supit{i}Harvard-Smithsonian Center for Astrophysics, 60 Garden St, Cambridge, MA 02138, USA\\
}

\authorinfo{Further author information: (Send correspondence to J.W.A.)\\J.W.A: E-mail: jappel3@jhu.edu}

\begin{document} 
  \maketitle 

\begin{abstract}
The Cosmology Large Angular Scale Surveyor (CLASS) experiment aims to map the polarization of the Cosmic Microwave Background (CMB) at angular scales larger than a few degrees. Operating from Cerro Toco in the Atacama Desert of Chile, it will observe over 65\% of the sky at 38, 93, 148, and 217~GHz.  In this paper we discuss the design, construction, and characterization of the CLASS 38~GHz detector focal plane, the first ever Q-band bolometric polarimeter array. 


\end{abstract}



\section{INTRODUCTION}
\label{sec:intro}  
The cosmic microwave background (CMB) radiation consists of photons emitted $\sim$13.8 billion years ago.  
Photons and charged particles in the early Universe were tightly coupled in a hot, dense plasma; as the universe expanded and cooled, the free electrons and protons combined to form neutral hydrogen, allowing the photons to decouple from matter and free stream throughout the Universe.
Today, we observe this radiation in all directions as an almost perfect \SI{2.725}{\kelvin} blackbody \cite{cobe_firas}. 
The near isotropy of the CMB poses a pressing question, how can two regions of the Universe that were not in causal contact at decoupling yield the same CMB radiation temperature to an observer today? 

The theory of Inflation\cite{guth_inflation,linde} explains our isotropic and homogeneous Universe by postulating a period of exponential expansion that took place in the first fraction of a second, smoothing inhomogeneities to scales much larger than the observable Universe.
Only small inhomogeneities sourced from quantum fluctuations are left once inflation ends. 
These quantum fluctuations are believed to produce the nearly scale-invariant spectrum of perturbations observed in the CMB anisotropies. 
Inflation theory predicts both scalar (over- and under-densities) and tensor perturbations (gravitational waves) to the space-time metric. 
These perturbations generate both temperature and polarization anisotropy in the CMB. 
The CMB polarization anisotropy can be decomposed into E-mode and B-mode rank-2 tensor fields. 
While both scalar and tensor perturbations contribute to the E-mode signal, only tensor perturbations contribute to the B-modes \cite{kamionkowski,seljak_zaldariaga}.
Detection of B-modes provides one of the cleanest tests of Inflation.
Nevertheless it is a challenging task given an expected signal amplitude hundreds of times smaller than the CMB temperature anisotropies.

The small primordial B-mode signal is easily obscured by multiple foregrounds. 
At angular scales smaller than half a degree, the B-mode signal is dominated by gravitational lensing of E-modes into B-modes \cite{sptpol_lensing,polarbear_lensing}. 
Galactic synchrotron emission dominate the sky's polarized signal at frequencies below $\sim$\SI{65}{\giga\hertz}\cite{wmap_pol_foreground2007}, while polarized dust dominates at higher frequencies. 
The amount of Galactic foreground contamination depends strongly on the Galactic latitude of the observation. 
One strategy is to observe regions of expected low foreground emission, in the hope that the primordial B-mode signal dominates; another is to measure the sky polarization at multiple frequencies, and use the spectral information to subtract the foreground contribution. 
Experiments that have measured or are aiming to measure the CMB polarization signal include QUAD \cite{quad}, CBI, \cite{cbi}, BOOMERANG \cite{boomerang}, DASI \cite{dasi}, WMAP \cite{wmap_9year}, CAPMAP \cite{capmap}, BICEP \cite{bicep}, QUIET \cite{quiet_2012}, BICEP2 \cite{bicep2_arxiv_exp}, KECK \cite{keck}, PLANCK \cite{planck_inter_pol}, POLARBEAR \cite{polarbear}, SPIDER \cite{spider}, EBEX \cite{ebex}, ABS \cite{abs_hwp}, ACTPOL \cite{actpol}, SPTPOL \cite{sptpol}, SPT3G \cite{spt3g}, GROUNDBIRD \cite{groundbird}, LITEBIRD \cite{litebird}, QUIJOTE \cite{quijote}, BICEP3 \cite{bicep3}, PIPER \cite{piper} and CLASS \cite{class_aas}.
In March 2014 the BICEP2 team \cite{bicep2_arxiv_bmode}, claimed a detection of primordial B-modes consistent with a tensor-to-scalar ratio $r$ of 0.2, in tension with $r<0.11$ \cite{planck_inflation} upper limit from temperature anisotropy measurements.
Confirmation of this claim requires better understanding of the polarized foregrounds in the BICEP2 patch, information that Planck data may provide in the near future.
Also critical are follow-up measurements from multiple experiments targeting the same and/or other sky regions with different sets of systematic issues.

Going forward, an ideal B-mode measurement would strive for high-sensitivity full-sky polarization maps at multiple frequencies.
Multi-frequency observations are required for foreground subtraction.
Full sky maps would provide access to the lowest multipoles, which are expected to contain an enhancement in the B-mode power spectrum generated from primordial gravitational waves entering the horizon during the reionization epoch.
The Cosmology Large Angular Scale Surveyor (CLASS) is designed to achieve these survey goals.
From the Atacama desert of Chile it will map over 65\% of the sky (45\% after masking the Milky Way) at 38, 93, 148, and 217~\SI{}{\giga\hertz}, employing Transition Edge Sensor (TES) bolometer arrays with $\SI{}{\micro\kelvin\sqrt{\second}}$ sensitivity. 
Each frequency band has a Variable-delay Polarizer Modulator (VPM) that converts the detector's polarization sensitivity between Stokes parameters Q and V. 
This modulation allows the signal band to be placed at $\sim\SI{10}{\Hz}$, away from instrumental and atmospheric 1/f noise.
For a detailed description of the CLASS project see [\citenum{tom_spie}] in these proceedings.

\begin{table}[h]
\caption{CLASS Q-band survey characteristics }
\label{tab:qband}
\begin{center}
\begin{tabular}{ll} 
\hline
\rule[-1ex]{0pt}{3.5ex}  Frequency & \SI{33}{\giga\hertz} to \SI{43}{\giga\hertz}  \\
\rule[-1ex]{0pt}{3.5ex}  Beam FWHM & \SI{1.5}{\degree}   \\
\rule[-1ex]{0pt}{3.5ex}  Field of view & $\SI{14}{\degree} \times \SI{19}{\degree}$   \\
\rule[-1ex]{0pt}{3.5ex}  Number of polarimeters& $36$   \\
\rule[-1ex]{0pt}{3.5ex}  Observing site & Cerro Toco in the Atacama desert   \\
\rule[-1ex]{0pt}{3.5ex}  Fraction of sky observed & 65\% \\ 
\rule[-1ex]{0pt}{3.5ex}  Polarization Modulation & VPM  \\
\rule[-1ex]{0pt}{3.5ex}  Detectors & Feedhorn-coupled Planar OMT TES bolometers  \\
\rule[-1ex]{0pt}{3.5ex}  Cryogenics & Dilution fridge with $T_{min}=\SI{30}{\milli\kelvin}$  \\
\rule[-1ex]{0pt}{3.5ex}  Mount & Azimuth, elevation, and boresight axis.  \\
\hline
\end{tabular}
\end{center}
\end{table}

In the following sections we present details of the design and construction of the CLASS \SI{38}{\giga\hertz} (Q-band) camera.
See Table \ref{tab:qband} for a summary of the CLASS Q-band survey characteristics.
This article is divided as follows: Section \ref{sec:opt} describes the focal plane optics and mechanical layout.
Section \ref{sec:tes_bol} discusses the TES bolometer design, the detector SQUID readout, and estimates of the final array sensitivity.
Section \ref{sec:conclusions} summarizes the camera's current status and future path.

\section{FOCAL PLANE OPTICS AND MECHANICAL DESIGN}
\label{sec:opt}

The CLASS Q-band focal plane (FP) mounts through copper rods onto the \SI{100}{\milli\kelvin} stage of a Bluefors\footnote{Bluefors Cryogenics, Arinatie 10, 00370 Helsinki, Finland +358 9-2245110} dilution refrigerator, which provides in excess of \SI{300}{\micro\watt} of cooling power at this temperature.
A low FP operating temperature is required by the low noise transition edge sensor (TES) bolometers described in the following section.
A \SI{24}{\centi\meter} by \SI{33}{\centi\meter} gold-plated copper monolithic FP baseplate minimizes thermal gradients.
It contains 36 detector-mounting locations, each with a light-coupling waveguide machined through the baseplate.
The cross-section of these waveguides transitions in one step from a circle of \SI{3.14}{\milli\meter} radius to a square with \SI{5.7}{\milli\meter} side length and \SI{1}{\milli\meter} corner radius.
Space on the edge of the baseplate is reserved for mounting the cold readout electronics.

\begin{figure}
\begin{center}
\begin{tabular}{c}
\includegraphics[width=0.3\paperwidth]{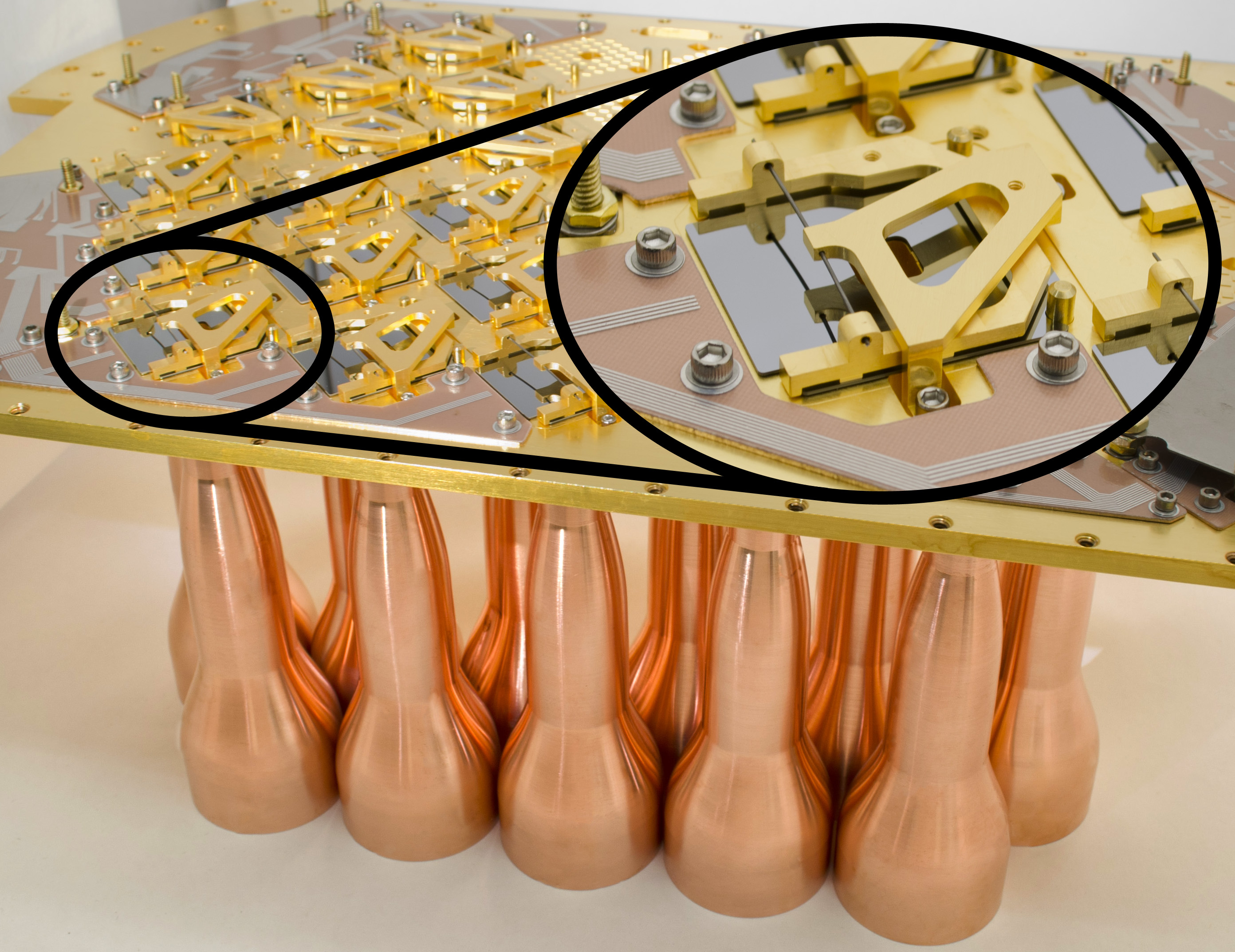}
\includegraphics[width=0.43\paperwidth,trim=90 220 0 18, clip]{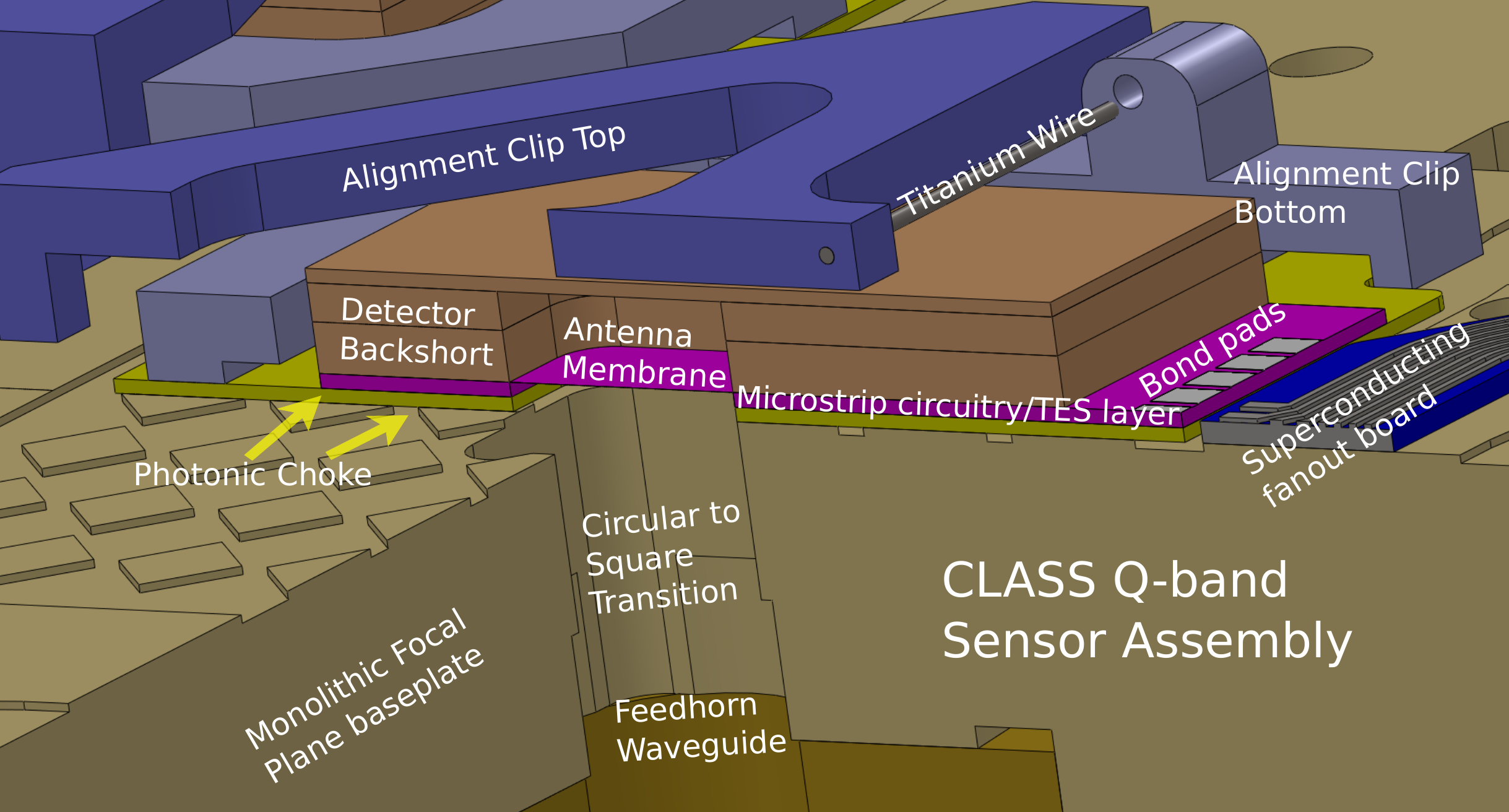}
\end{tabular}
\end{center}
\caption{The picture on the left shows the CLASS Q-band focal plane with feedhorns, Al circuit boards, mounting jigs, and choke chips installed.
The magnified section shows a choke chip aligned by the two-piece spring-loaded jig.
The 3D model on the right depicts a detector pixel mounted on the focal plane baseplate.
The backshort cap (brown), detector chip (fuchsia), and interface chip (yellow) are hybridized and mounted as a single assembly.
In this view the detector and alignment jig are cut down the middle, while one quarter of the baseplate is removed, exposing the circular to square waveguide transition.
}
\label{fig:fp_pic_model}
\end{figure}

A smooth-walled copper feedhorn\cite{lingzhen,joseph_SPIE} 
 bolts to each waveguide on the baseplate, thus coupling light onto the detectors mounted on the opposite side.
The feedhorn's flange bolts to the baseplate aligning a cylindrical boss at the bottom of the feedhorn to a matching (slightly oversized) cylindrical extrusion on the baseplate. 
This setup aligns the feedhorn and baseplate waveguides to $\SI{25}{\micro\meter}$ tolerance.
A photonic choke joint\cite{ed_photonic_choke} is used to define the interface between the integrated silicon detector and the baseplate. 
On the baseplate side the interface consists of a 7x7 grid of square pillars machined into the plate.
The pillars are \SI{2.28}{\milli\meter} on a side, spaced \SI{3.88}{\milli\meter} apart and rotated \SI{45}{\degree} with respect to the detector waveguide.
The pillar height is set to $0.18\pm\SI{0.01}{\milli\meter}$.
Figure \ref{fig:fp_pic_model} shows a model of the feedhorn-baseplate-detector stack.

Thirteen detector chips are fabricated on one \SI{100}{\milli\meter} silicon wafer\cite{detector_fab}\nocite{crowe_nasa_fab}.
Each detector is then diced and attached to a metalized silicon ``choke" chip that provides the other half of the photonic choke joint.
Cooling the focal plane to \SI{100}{\milli\kelvin} shrinks the copper baseplate by 0.33\%.
On the other hand, the silicon chips contract by a negligible amount.
The baseplate's contraction is compensated for in the machining dimensions, while the differential contraction with respect to the detectors is solved with a mounting scheme that accurately aligns the detector pixel to the waveguides without over-constraining the silicon.
The mounting scheme shown in Figure \ref{fig:fp_pic_model} employs two pins located on the baseplate ($\SI{10}{\micro\meter}$ tolerance) to constrain the pixel's displacement and rotation. 
A spring-loaded jig applies $\sim\SI{0.1}{\newton}$ on the pixel ($\sim20\times$ its weight) in the direction of the pins and downward toward the baseplate.
This jig is composed of two pieces called the ``clip" and the ``tensioner."
The clip bolts to the baseplate, while the tensioner is attached to the clip via a titanium alloy (Ti-6AL-4V) wire that acts as the spring.
A ``V"-shape feature on each pixel aligns with one of the pins constraining its location on the baseplate, while a flat edge on the other side of the pixel is registered by the second pin to constrain rotation. 

The dominant TE$_{01}$ and TE$_{10}$ modes of the square waveguide are separated by a planar orthomode transducer (OMT) \cite{wollack_omt,class_omt,karwan_spie_2014} that utilizes magic-T\cite{magicT} and via-less crossover\cite{crossover} designs.
The symmetry in the OMT circuit enables broadband operation over the 2:1 waveguide bandwidth.
After the OMT, a reactive filter defines the desired bandwidth of 33-\SI{43}{\giga\hertz} for the CLASS Q-band focal plane.
This bandpass range is optimized for maximum signal-to-noise based on the atmospheric emission at the CLASS observing site in the Atacama desert.
To minimize dielectric loss, the OMT is fabricated on \SI{5}{\micro\meter} single-crystal silicon\cite{detector_fab}, which serves as the low-loss dielectric as well as the support membrane for the TES bolometers\cite{karwan_spie_2014}.
Preliminary measurements of the optical efficiency of the entire polarimeter, from OMT to TES bolometers, is 90\% in each polarization\cite{karwan_spie_2014}.

Optical power coupling through each polarization channel (2 channels per pixel, 72 total in the Q-band array) is dissipated on a matched resistive element that is thermally connected to a TES bolometer.
Individual bolometers are voltage-biased by a $\sim$$\SI{250}{\micro\ohm}$ shunt resistor and noise bandwidth limited by a Nyquist inductor of $\sim$$\SI{310}{\nano\henry}$.
Both components are located on the ``interface chip'' fabricated by NIST, and mounted at the edge of the focal plane.
Connections between the detectors and the interface chip are composed of superconducting aluminum (Al) bonds to aluminum traces fabricated on silicon and on FR4 substrates.
Aluminum silicon circuits are fabricated in-house and then laser-diced, while \SI{0.25}{\milli\meter} thick aluminum FR4 circuits backed by \SI{0.51}{\milli\meter} thick copper surfaces for good thermal conductivity are fabricated by Tech-etch\footnote{Tech-etch, 45 Aldrin Road, Plymouth, MA 02360, (508)747-0300}.
Trace lengths between the detectors and the shunt resistors range from \SI{2}{\centi\meter} to \SI{15}{\centi\meter}, increasing the inductance of the TES loop by as much as \SI{200}{\nano\henry}.

The TES bolometers are read out by superconducting quantum interference devices\cite{nist} (SQUID) mounted next to the interface chips, and connected via Al bonds.
Eight SQUID multiplexing chips are mounted on the FP, sandwiched between \SI{0.51}{\milli\meter} thick Niobium (Nb) sheets for magnetic shielding.
Each chip may read out 11 channels for a total of 88, divided as follows: 72 optical bolometers, 4 optically-isolated dark bolometers for detecting light leaks and monitoring the stability of the bath temperature, and 12 dark SQUIDS to monitor readout noise and magnetic pick-up. 
SQUID channels on the FP are connected to a \SI{4}{\kelvin} stage of SQUID series array (SA) amplifiers via NbTi superconducting cables and multi-layer copper circuit boards with MDM\footnote{Glenair, 1211 Air Way, Glendale, CA 91201-2497, (818)247-6000} connectors.
A \SI{4}{\kelvin} circuit board holds the SA and serves as interface between the NbTi cable launched towards the FP, and two low-thermal-conductivity Manganin MDM cables that carry signal and bias lines to the warm readout electronics.
All wires are twisted pair to reduce pick-up; MDM connectors are strain-relieved with low-profile epoxy back-shells; and wires are wound with aramid fiber for durability. 
The NbTi and manganin cables are fabricated by Tekdata cryoconnect\footnote{Tekdata Interconnections Limited, Festival Way Stoke-on-Trent, Staffordshire, ST1 5SQ, UK, +44 (0) 1782-254-700}.
The warm readout Multi-Channel Electronics (MCE) designed for Time Domain Multiplexing (TDM) of SQUIDS is provided by the University of British Columbia (UBC). 
Similar systems are operated successfully in the field by multiple CMB experiments \cite{actpol,abs_hwp,bicep2_arxiv_exp,spider}.


\section{TES BOLOMETERS} 
\label{sec:tes_bol}

TES bolometers measure the optical power that couples through the planar OMT antennas of the Q-band detector pixels.
These TESs are based on a MoAu superconducting bilayer with a critical temperature ($T_c$) target of $\sim\SI{150}{\milli\kelvin}$, a normal resistance ($R_n$) of $\sim\SI{10}{\milli\ohm}$ and a typical operating resistance ($R_{tes}$) of $\sim\SI{5}{\milli\ohm}$\cite{detector_fab}.
The bilayer sits on an island that is thermally isolated from the supporting structure through silicon legs of thermal conductivity $G$.
It is voltage-biased by a shunt resistor ($R_{sh}$), such that changes in TES resistance may be measured as a current signal through a SQUID connected in series with the TES.
For example, consider an optical signal that deposits power on the TES island through the OMT's microstrip circuitry.
This signal raises the island's temperature, and hence also the resistance of the bilayer.
This change in resistance is converted to a change in current by the voltage bias circuit, which is then read-out by the SQUID amplifier.
The following sections discuss the optimization of the TES bolometer design for sensitivity, stability, and uniformity across the array.

\subsection{TES Design Choices, and Results}
\label{sec:tes_choice}

\begin{figure}
\begin{center}
\includegraphics[width=0.8\paperwidth]{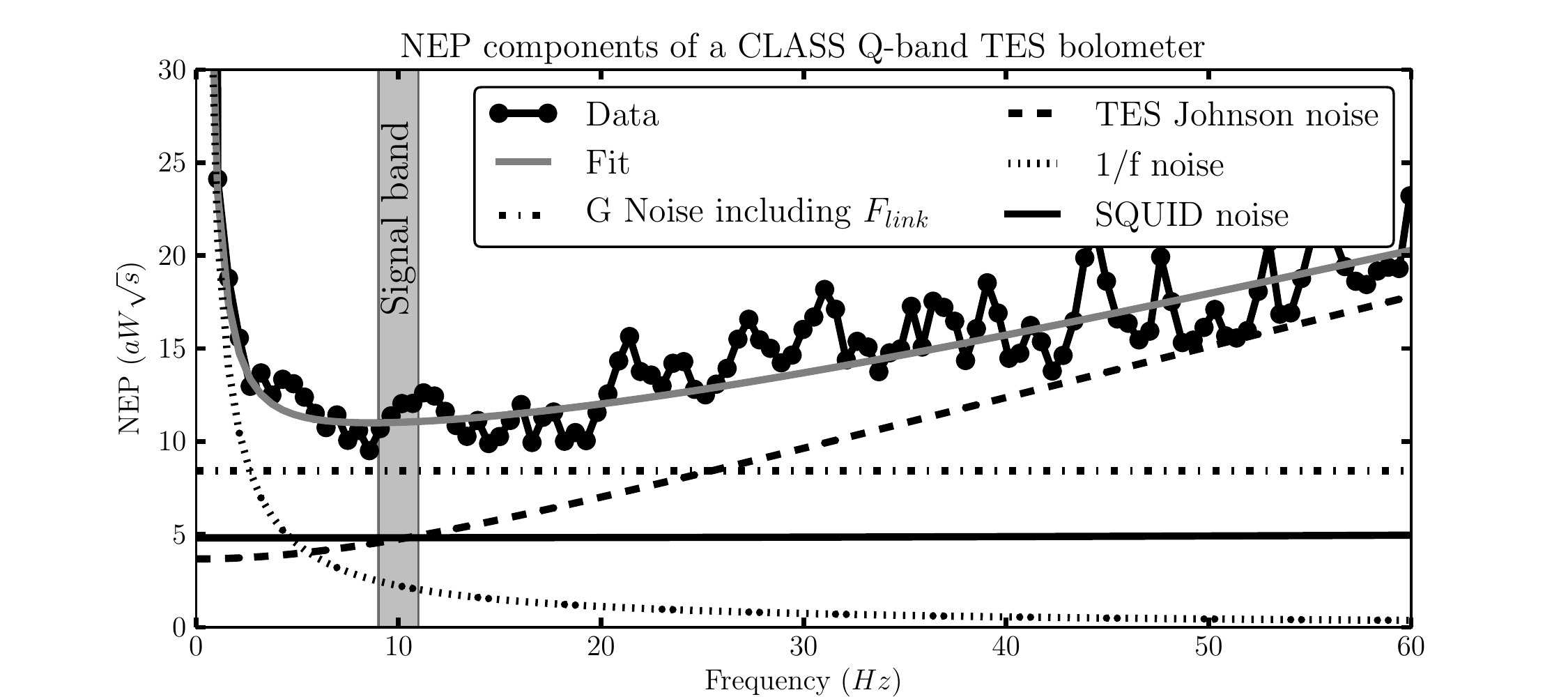}
\caption{ NEP components of CLASS Q-band pixel operated at $T_b=\SI{70}{\milli\kelvin}$ with a backshort cover.
The expected NEP from thermal fluctuations is $\SI{8.5}{\atto\watt\sqrt{\second}}$, but measurements imply total detector NEP of $\SI{11}{\atto\watt\sqrt{\second}}$ at 10~Hz.
This difference is accounted for through three components: SQUID readout noise, detector Johnson noise, and $1/f$ noise.
In the data above, the detectors were read-out by a mux09 SQUID multiplexing chip, in the final focal plane mux11d chips will be used at a higher multiplexing rate, which provides a factor of root two reduction in SQUID NEP (see section \ref{sec:squid_readout}).
Detector Johnson noise is suppressed through electro-thermal feedback (ETF) at lower frequencies, but increases lower on the transition as $\beta$ gets larger \cite{irwin,irwin_nonequilibrium}. 
Optimal ETF requires TES bias power greater than the expected optical loading, hence the requirement $P_{sat}=P_{load}+P_{bias} \geq 2 P_{load}$.
In the data above no temperature servo was applied, therefore improved $T_b$ stability should further reduce the $1/f$ noise component.
}
\label{nep_Q6}
\end{center}
\end{figure}

The thermal fluctuation or phonon noise of a TES bolometer is given by:
\begin{equation}
\mbox{NEP}_G=\sqrt{2 k_b T_c^2 G F_{link}},
\label{nep_eqn}
\end{equation}
\begin{equation}
F_{link}=\frac{1+(T_b/T_c)^{n+1}}{2},
\label{flink_eqn}
\end{equation}
where NEP$_G$ is the phonon noise equivalent power with units \SI{}{\watt\sqrt{\second}}, $T_{b}$ is the temperature of the heat bath supporting the TES legs, $n$ is the exponent governing power flow between the TES and the heat bath, and $k_b$ is Boltzmann's constant. 
$F_{link}$ depends on the nature of the energy transport between the TES and the bath.
Equation \ref{flink_eqn} is valid in the ballistic limit when the mean free path of the phonons is larger than the length of the thermal link\cite{boyle_flink}.
It applies to the CLASS bolometers, for which the thermal conductivity is dominated by a short silicon leg \cite{leg_precision}.  

The detector's saturation power ($P_{sat}$) is defined as the amount of power required to raise the island temperature to $T_c$ and can be computed from:
\begin{equation}
    P_{sat}=\kappa(T_c^n-T_{b}^n),
\label{psat_eqn}
\end{equation}
where $\kappa$ is a constant given by the geometry of the TES legs, and $n$ is equal to four in the limit of ballistic phonons.
$P_{sat}$ is equivalent to the largest signal a TES bolometer can measure.

    The expected background optical loading in the field for the CLASS Q-band camera is $\SI{1.7}{\pico\watt}$ \cite{tom_spie}, while the expected background optical noise is $\SI{12}{\atto\watt\sqrt{\second}}$ \cite{tom_spie}.
To achieve background-noise-limited detectors, the CLASS Q-band camera requires TES bolometers with low $T_c$ and $G$ values.
On the other hand, the bolometer's $P_{sat}$ has to be larger than the expected optical loading (typically a factor of two or more for optimal performance), hence this limits how low $G$ and $T_c$ can be, given a target operating $T_{b}$.
$G$ is related to the $P_{sat}$ parameters through:
\begin{equation}
G=\left.\frac{dP_{sat}}{dT}\right|_{T_c}=n \kappa T_c^{n-1}.
\label{g_eqn}
\end{equation}
Equation \ref{nep_eqn} can be expressed in terms of the target $P_{sat}$, $T_{b}$, and $T_c$ through equations \ref{psat_eqn} and \ref{g_eqn}.
NEP$_{G}$ is then minimized for an optimal $T_c$ ($T_c^{opt}$) given a target $T_{b}$:
\begin{equation}
T_c^{opt}= 1.65 \times T_{b}~\mbox{when}~n=4.
\end{equation}
Once the optimal $T_c$ target is known, $\kappa$ is chosen to satisfy the $P_{sat} \geq 2 P_{load}$ condition, where $P_{load}$ is the expected optical loading.

Tests of the first CLASS dilution refrigerator without the cryostat window reach a minimum $T_{b}$ of $\SI{27}{\milli\kelvin}$.
Including the expected optical loading through the window implies a $T_{b}$ as low as \SI{50}{\milli\kelvin}.
The $T_c$ targets for the Q-band detector wafers were conservatively chosen based on a $T_{b}$ target of $\SI{90}{\milli\kelvin}$, which implies a $T_c^{opt}$ target of $\sim$~$\SI{150}{\milli\kelvin}$.
The average $T_c$ of the science-grade detectors fabricated so far is $\SI{156}{\milli\kelvin}$.

CLASS TES detectors have been fabricated and tested with multiple leg designs\cite{leg_precision}.
The optimal design for the $\SI{38}{\giga\hertz}$ detectors contains a \SI{10}{\micro\meter} long, \SI{12}{\micro\meter} wide leg that dominates the thermal conductance to the island.
This design has yielded detectors with an average $\kappa=\SI{12.1}{\nano\watt\per\kelvin^n}$  and a standard deviation across a detector wafer of $\sim$5\%.
This average $\kappa$ value implies a Q-band $P_{sat}$ of \SI{6.8}{\pico\watt} and NEP$_{G}$ of \SI{8}{\atto\watt\sqrt{\second}} at $T_{b}=\SI{70}{\milli\kelvin}$ and $T_c=\SI{156}{\milli\kelvin}$. 
This scenario provides an attractive ratio of $P_{sat}/P_{load}=4$, which would allow some excess loading and variation in the detector parameters without great detriment to the detector performance.

Detector wafers consistently yield devices with $T_c$ values between \SI{140}{\milli\kelvin} and \SI{165}{\milli\kelvin}.
Furthermore the standard deviation of individual detector $T_c$ within a wafer is only $\sim \SI{3.4}{\milli\kelvin}$ (see Figure \ref{hist_deltaTc}).  
The total dark NEP of a bolometer on a fully assembled pixel has been measured directly from the power spectra of time-ordered data acquired during dark tests, achieving values as low as $\SI{11}{\atto\watt\sqrt{\second}}$ (see Figure \ref{nep_Q6}).
This value is consistent with the detector's parameters ($T_c=\SI{162}{\milli\kelvin}$, $G=\SI{193}{\pico\watt}$) when including the SQUID readout and detector Johnson noise.
Detectors from lower $T_c$ wafers are expected to have detector NEP in the field of $\SI{9}{\atto\watt\sqrt{\second}}$, excluding photon noise. 

\subsection{TES stability}

\begin{figure}
\begin{center}
\includegraphics[width=0.8\paperwidth]{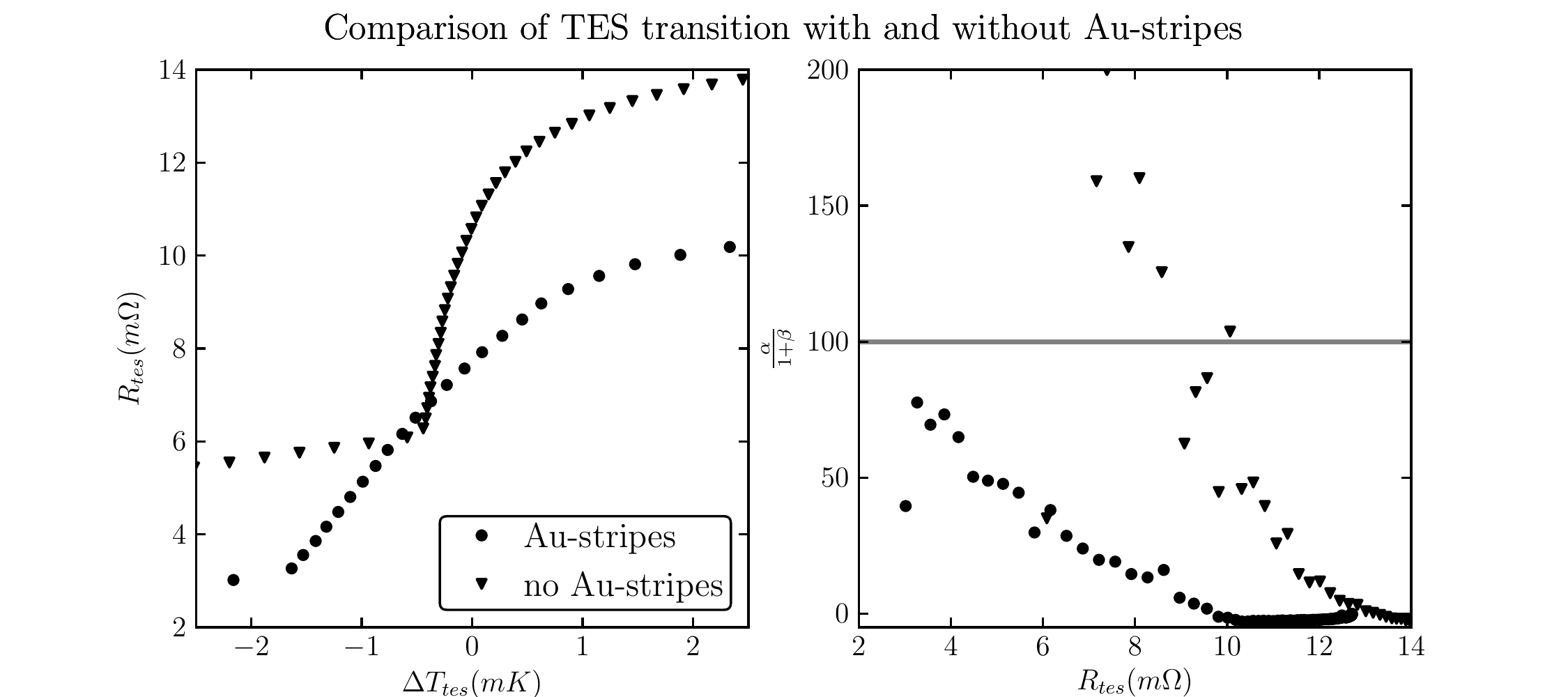}
\caption{ On the left is a plot of the TES T vs R transition (extracted from I-V curves) for bilayers with and without gold Au-stripes.
Note that the transitions do not reach zero resistance because the TES becomes unstable at lower resistance, due partly to the high $L$ value used during these measurements.
The right plot shows the parameter $\alpha/(1+\beta)$ across the transition (estimated from I-V curves) for the two types of TES detectors. 
The TES with stripes has a gentler R-T curve slope and lower $\alpha/(1+\beta)$ throughout the transition, hence adding stripes improves the stability of the detector.
}
\label{IV_stability}
\end{center}
\end{figure}

In the simplest TES model there are two time constants: $L/R_{tes}$ and $\frac{C/G}{1+\mathcal{L}_I/(1+\beta)}$.
$L$ is the inductance of the TES loop, $C$ the heat capacity of the TES, $\beta=\frac{I_{tes}}{R_{tes}}\frac{dR_{tes}}{dI_{tes}}$ and $\mathcal{L}_I$ the loop gain of the negative electro-thermal feedback (ETF) loop created by voltage-biasing the TES. $\mathcal{L}_I$ can be calculated  from:
\begin{equation}
\mathcal{L}_I=\frac{P_{tes}\alpha}{G T_c},
\label{Lgain}
\end{equation}
where alpha is equal to $\alpha=\frac{T_{tes}}{R_{tes}}\frac{dR_{tes}}{dT_{tes}}$.

If the electrical time constant ($L/R_{tes}$) becomes comparable to the thermal time constant ($\frac{C/G}{1+\mathcal{L}_I/(1+\beta)}$) then the TES becomes unstable and oscillates, making the detector inoperable.
The inductance of the TES loop ($L_{tes}$) is approximately $\SI{500}{\nano H}$, dominated by a Nyquist inductor included in the interface chip to limit the detector noise bandwidth, such that it can be sampled at \SI{20}{\kilo\hertz} without aliasing more than 2\% of the noise in the signal band.
Assuming $R_{sh} \ll R_{tes}$ , and $\mathcal{L}_{I} \gg 1$,  the stability condition of the TES given in [\citenum{irwin}] is equivalent to:
\begin{equation}
    \frac{L}{R_{tes}} < \frac{T_c C (1+\beta)}{P_{tes} \alpha},
\end{equation}
    where $P_{tes}=P_{sat}-P_{load}$.
This implies that a TES detector cannot operate low on the transition where $R_{tes}$ is small, unless $C$ is large enough and $\alpha/(1+\beta)$ is small enough for the target $T_c$ and $P_{tes}$. 

The $\SI{2.5}{\pico\joule\per\kelvin}$ heat capacity of the TES island is dominated by a $\SI{400}{\nano\meter}$ palladium layer\cite{leg_precision}.
By adding three gold bars (Au-stripes) that partially span the bilayer perpendicular to the direction of the bias current\cite{irwin,staguhn_nasa_nbars,sadler_prox_effect}, $\alpha/(1+\beta)$ is suppressed below 100 (the estimated threshold for stability) across the TES transition.
For comparison, a TES design without Au-stripes broke the $\alpha/(1+\beta) > 100$ stability mark around 60\% $R_n$ (see Figure \ref{IV_stability}).
Lowering $\alpha/(1+\beta)$ translates into stable detectors over a wider range of the TES transition, an important factor when multiple detectors share one bias line, and are typically biased at different points on the transition.
It is also expected to reduce non-equilibrium TES noise that is correlated to high $\alpha$ devices \cite{Ullom_excess}.

\subsection{TES Uniformity}

\begin{figure}
\begin{center}
\includegraphics[width=0.8\paperwidth]{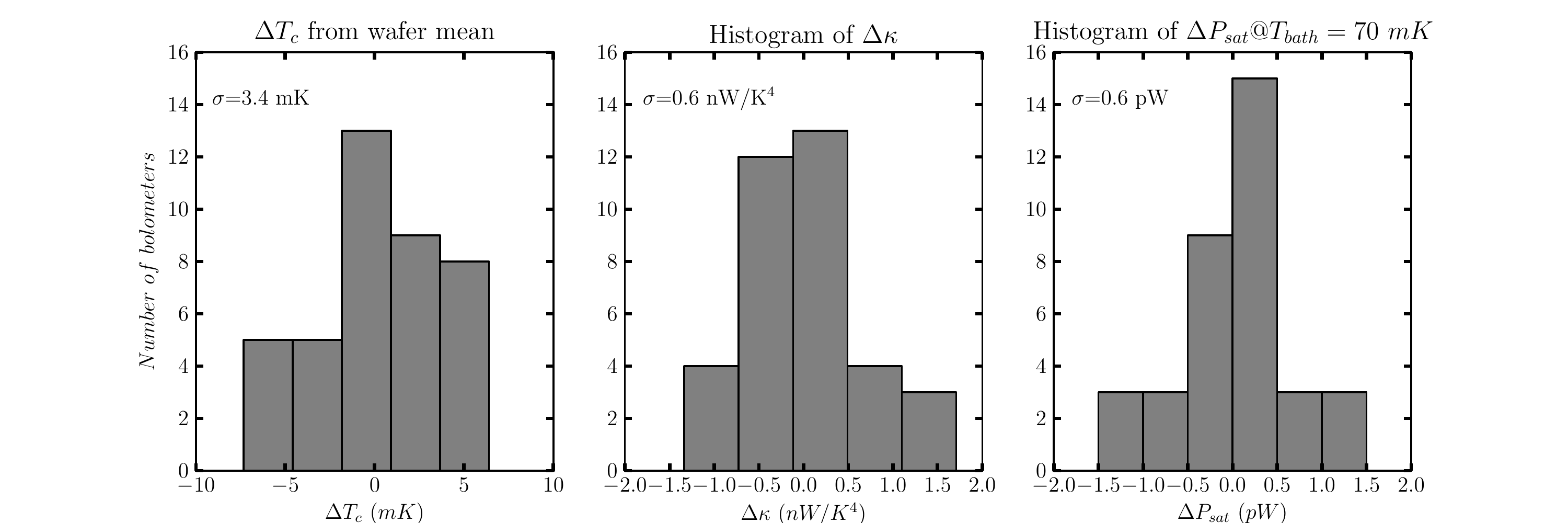}
\caption{ 
From left to right are histograms of $\Delta T_c$, $\Delta\kappa$, and $\Delta P_{sat}$ respectively, where $\Delta$ denotes the difference of the individual detector parameter value to the average within its wafer.
The histograms include measurements of science grade detectors from four Q-band CLASS wafers.
Each MCE detector bias line will group bolometers from the same wafer, hence good wafer uniformity translates to optimal TES biasing.
}
\label{hist_deltaTc}
\end{center}
\end{figure}

So far we have described tuning of individual detector parameters to optimize performance; equally important to maximizing the final instrument sensitivity is the uniformity of the detector fabrication.
Good wafer uniformity is required to achieve the optimal TES parameter targets on all detectors, or at the very least necessary to produce detectors that can operate at the same bias voltage, optical loading, and bath temperature.

The Q-band camera will run eight detector bias line groupings of up to 11 TESs each.
The $T_c$ uniformity within a Q-band wafer is shown in Figure \ref{hist_deltaTc}.
The standard deviation of $T_c$  within a science-grade wafer is only $\SI{3.4}{\milli\kelvin}$, while average $T_c$ of these wafers falls in the \SI{140}{\milli\kelvin} to \SI{165}{\milli\kelvin} range.

The uniformity of the thermal conductance is probed by measuring the thermal conductivity parameter $\kappa$ with I-V curves at multiple $T_b$ temperatures. 
Thirty-six such measurements are included in the second histogram of Figure \ref{hist_deltaTc} yielding a $\kappa$ standard deviation across detectors of $\sim$5\%\cite{leg_precision}. 
This small fluctuation in $\kappa$ is sub-dominant to the $T_c$ fluctuation, which is amplified by a factor of $4$ when considering its effect on $P_{sat}$.
By grouping the detectors in $T_c$ ranges of \SI{3}{\milli\kelvin} while discarding outliers, the effects on $P_{sat}$ due to $T_c$ variations become comparable to those from the $\kappa$ distribution. 

The standard deviation of $P_{sat}$ is less than 10\% across a wafer and should be lower within a bias group.
At this level the $P_{sat}$ disparities are comparable to those expected from differences in optical loading across the focal plane due to coupling efficiency variations, or elevation-dependent atmospheric loading.
The ability to group similar TES bolometers will allow for fine-tuning of the detector bias voltage based on observing conditions. The optimal bias point of operation for these detectors is found low on the transition where ETF strongly reduces the TES Johnson noise, but not so low that they become unstable. 
See Table \ref{tab:tes_param} for a summary of the CLASS Q-band TES bolometer parameters.

\subsection{SQUID readout}
\label{sec:squid_readout}

Low-noise current amplifiers are necessary ($\sim\SI{30}{\pico\ampere\sqrt{\second}}$) to readout the CLASS TES bolometers.
A single-channel SQUID  would be able to achieve this task with ease, but since the array holds many detectors, TDM is used to reduce the number of wires reaching the cold stages.
TDM increases the SQUID readout noise through aliasing by a factor that scales with the number of multiplexed channels and the time spent reading each.
CLASS utilizes the latest generation of 11 channel flux-activated\cite{SQUID_switch} mux11d\cite{code_flux_switches_nist} chips fabricated by NIST, which can be multiplexed faster than the earlier mux09 version.
The mux11d readout only uses two stages of SQUIDs, a stage one (SQ1) to readout the detector signal, and a $\SI{4}{\kelvin}$ series array (SA) of SQUIDs for further amplification before the warm readout electronics.
The previous mux09 version used a second stage SQUID (SQ2) in between the SQ1 and SA to aid with the multiplexing scheme.
The SQ2 voltage bias circuit typically limited the multiplexing speed of the system.

The MCE readout electronics runs on a \SI{50}{\mega\hertz} clock, with the Q-band mux11d readout set to wait as little as 50 clock cycles per row switch, and acquire between 10 and 50 clock cycles of data before the next row switch. 
Currently the plan is to run a 50-50 cycle of wait-sample for each channel. 
This translates to a SQUID noise aliasing factor of 22, with an individual detector sampling frequency of \SI{45}{\kilo\hertz}, large enough to avoid aliasing of high-frequency detector noise.
Initial noise measurements of the Q-band mux11d readout place its amplitude at $\sim\SI{20}{\pico\ampere\sqrt{\second}}$. 
A typical detector responsivity of $\SI{150}{\nano\volt}$ then implies a SQUID NEP of $\SI{3}{\atto\watt\sqrt{\second}}$.
Note that it adds in quadrature to the detector NEP ($\sim\SI{11}{\atto\watt\sqrt{\second}}$) and photon NEP ($\sim\SI{12}{\atto\watt\sqrt{\second}}$).
Hence the SQUID readout only contributes a few percent to the total noise.

\subsection{Sensitivity projections}
\label{sec:sens_proj}

The total detector NEP ($\mbox{NEP}_{tot}$) during observations can be estimated by adding (in quadrature) the nominal detector dark NEP ($\mbox{NEP}_{dark}$) measured in section \ref{sec:tes_choice} ($\sim\SI{11}{\atto\watt\sqrt{\second}}$) to the expected NEP contribution from optical loading ($\mbox{NEP}_{\gamma}$) due to the intrinsic telescope emissions, the atmosphere, and the CMB. 
The total optical loading ($P_{opt}$) on the detector during observations is expected to be $\SI{1.7}{\pico\watt}$, generating an $\mbox{NEP}_{\gamma}\sim\SI{12}{\atto\watt\sqrt{\second}}$.
We expect $69\%$ coupling efficiency to the sky signal ($\epsilon$) , while the conversion factor from sky power to CMB temperature ($\frac{dP}{dT_{cmb}}$) is equal to \SI{0.13}{\pico\watt\per\kelvin}.
For more details on the optical loading and efficiency estimates see~[\citenum{tom_spie}].
A noise-equivalent CMB temperature ($\mbox{NET}_{cmb}$) of $\SI{181}{\micro\kelvin\sqrt{\second}}$ is obtained for a single Q-band bolometer from:
\begin{equation}
\mbox{NET}_{cmb}=\frac{\mbox{NEP}_{tot}}{\epsilon} \frac{dT_{cmb}}{dP}.
\end{equation}
Depending on the detail of the VPM modulation, the time spent measuring Stokes parameter Q or V can be split 50/50 ($\epsilon_Q=0.5$) or possibly up to 85/15 ($\epsilon_Q=0.85$). 
This translates to noise equivalent Q ($\mbox{NEQ}$) range from $\SI{363}{\micro\kelvin\sqrt{\second}}$ to $\SI{214}{\micro\kelvin\sqrt{\second}}$.
Assuming $85\%$ of the time is spent measuring Q and $\sim 90\%$ of the detectors operating, then the expected total array NEQ is $\SI{27}{\micro\kelvin\sqrt{\second}}$.
This array NEQ means CLASS can map 65\% of the sky to WMAP Q-band sensitivity ($\SI{280}{\micro\kelvin~arcmin}$\cite{wmap_9year}) with 249 hours of observations, and to the expected final Planck Q-band sensitivity ($\SI{142}{\micro\kelvin~arcmin}$\cite{planck_bluebook}) in 969 hours.
The low emissivity of the atmosphere at Q-band even for relatively high precipitable water vapor (PWV), combined with the continuous dilution refrigerator cycle should permit an observing efficiency from the Atacama similar to the 60\% efficiency achieved by the QUIET Q-band receiver \cite{quiet_qband}.
Therefore in a three-year survey CLASS can obtain $\sim$15000~hours of data, yielding a Q-band map sensitivity four times greater than Planck.


\begin{table}[h]
\caption{Q-band detector parameters}
\label{tab:tes_param}
\begin{center}
\begin{tabular}{|l|l|ccc|l|l|} 
\cline{1-2}
\cline{6-7}
\rule[-1ex]{0pt}{3.5ex} Parameter & Value & && & Parameter & Value \\ 
\cline{1-2}
\cline{6-7}
\rule[-1ex]{0pt}{3.5ex}  $T_c$ & \SI{156}{\milli\kelvin} & && & $C$  & \SI{2.5}{\pico\joule\per\kelvin} \\ 
\cline{1-2}
\cline{6-7}
\rule[-1ex]{0pt}{3.5ex}  $T_{b}$ &  \SI{70}{\milli\kelvin} & && &  $\mathcal{L}_I/(1+\beta)$ & 4-18 \\ 
\cline{1-2}
\cline{6-7}
\rule[-1ex]{0pt}{3.5ex}  $R_n$ & \SI{10}{\milli\ohm} & && & $F_{3db}$  & \SI{40}{\hertz}-\SI{190}{\hertz}  \\ 
\cline{1-2}
\cline{6-7}
\rule[-1ex]{0pt}{3.5ex}  $G@Tc$ &  \SI{184}{\pico\watt/\kelvin} & && & $\mbox{NEP}_{G}$ & $\SI{8}{\atto\watt\sqrt{\second}}$ \\ 
\cline{1-2}
\cline{6-7}
\rule[-1ex]{0pt}{3.5ex}  $\kappa$ & $\SI{12.1}{\nano\watt\per\kelvin^4}$ & && & $P_{load}$  & $\SI{1.7}{\pico\watt}$  \\ 
\cline{1-2}
\cline{6-7}
\rule[-1ex]{0pt}{3.5ex}  $n$ &  $4$  & && & $\mbox{NEP}_{tot}$  & $\SI{16}{\atto\watt\sqrt{\second}}$ \\ 
\cline{1-2}
\cline{6-7}
\rule[-1ex]{0pt}{3.5ex}  $P_{sat}$ & \SI{6.8}{\pico\watt} & && & $dP/dT_{cmb}$  &  \SI{0.13}{\pico\watt\per\kelvin} \\ 
\cline{1-2}
\cline{6-7}
\rule[-1ex]{0pt}{3.5ex}  $R_{sh}$ & \SI{250}{\micro\ohm} & && & $\epsilon$, $\epsilon_{Q}$  & 0.69, 0.85   \\ 
\cline{1-2}
\cline{6-7}
\rule[-1ex]{0pt}{3.5ex}  $L_{tes}$ & \SI{500}{\nano\henry} & && & $\mbox{NEQ}_{array}$  & $\SI{27}{\micro\kelvin\sqrt{\second}}$  \\ 
\cline{1-2}
\cline{6-7}
\end{tabular}
\end{center}
\end{table}

\section{CONCLUSIONS} 
\label{sec:conclusions}

The CLASS Q-band camera is the first-ever bolometric polarimeter array in its frequency band. 
The focal plane is constructed around a monolithic copper baseplate that serves as the interface between smooth-walled feedhorns and planar OMT circuitry coupled to TES bolometers.
The baseplate provides mechanical support for the superconducting wiring of the detector readout and includes a set of spring-loaded jigs used to mount and align individual silicon detector pixels.
We have validated the TES bolometer design yielding \SI{11}{\atto\watt\sqrt{\second}} dark detector NEP with $\SI{6.8}{\pico\watt}$ saturation power,  while maintaining stability across most of the transition. 
Detector $T_c$ and $\kappa$ statistics show a well-constrained fabrication process that has led to a uniform detector set, optimally biasable to achieve an array NEQ of $\sim$$\SI{27}{\micro\kelvin\sqrt{\second}}$.
Within a year of the 2015 observing start date, the CLASS Q-band array should provide the most sensitive polarization data at \SI{38}{\giga\hertz} across 65\% of the sky.
The Q-band camera will be the first deployed by the CLASS experiment, with the main goal of gathering information on low-frequency CMB foregrounds such as synchrotron. 
Its final Q-band maps are a critical element necessary to disentangle the primordial CMB B-mode signal from dominant foreground polarized emissions.

\acknowledgments     

The CLASS project is supported by the National Science Foundation under Grant Number 0959349.
The detectors used for CLASS have been developed under a NASA ROSES/APRA effort.
Detector development work at JHU was funded by NASA grant number NNX14AB76A.
CLASS operates in the Parque Astron\'{o}mico Atacama in northern Chile under the auspices of the Comisi\'{o}n Nacional de Investigaci\'{o}n Cient\'{i}fica y Tecnol\'{o}gica de Chile (CONICYT). 
T. E.-H. is supported by a National Science Foundation Astronomy and Astrophysics Postdoctoral Fellowship.

\def\jnl@style{\it}
\def\aaref@jnl#1{{\jnl@style#1}}
\def\apj{\aaref@jnl{ApJ}}
\def\prd{\aaref@jnl{Phys.~Rev.~D}}
\def\nat{\aaref@jnl{Nature}}
\def\apjs{\aaref@jnl{ApJS}}
\def\pasp{\aaref@jnl{PASP}} 
\def\aap{\aaref@jnl{A\&A}}
\bibliography{Qband}   
\bibliographystyle{spiebib}   

\end{document}